\documentclass[aps,pre,twocolumn,longbibliography,epsfig,floats,preprintnumbers,floatfix]{revtex4-2}
\usepackage{graphicx,amsmath,amssymb,bm,color,sidecap}
\usepackage[latin2]{inputenc}
\usepackage{dcolumn,nicefrac,blindtext,nicefrac,pgfplots}

\newcommand{\fijvec}{\vec f\!_{_{ij}}}
\newcommand{\hij}{h_{_{ij}}}

\newcommand{\ecijhat}{{\bm\hat{\bf\text{e}}}_{c_{_{ij}}}}

\newcommand{\rg}{r\!_{_\text{g}}}
\newcommand{\rgi}{r\!_{_{\text{g},i}}}
\newcommand{\fpi}{f\!_{_{\|i}}}

\begin{document}
\title{Mechanical Interactions Govern Self-Organized Ordering in Bacterial Colonies on Surfaces}
\author{Samaneh Rahbar}
\affiliation{Department of Theoretical Physics and Center for Biophysics, 
Saarland University, 66123 Saarbr\"ucken, Germany}
\author{Ludger Santen}
\affiliation{Department of Theoretical Physics and Center for Biophysics, 
Saarland University, 66123 Saarbr\"ucken, Germany}
\author{Reza Shaebani}
\email{shaebani@lusi.uni-sb.de}
\affiliation{Department of Theoretical Physics and Center for Biophysics, 
Saarland University, 66123 Saarbr\"ucken, Germany}

\begin{abstract}
Bacterial colonies growing on surfaces are shaped by mechanical stresses transmitted 
through the community, governed by the balance between cell growth and steric and 
cell-substrate interactions. Using overdamped dynamics simulations of nonmotile, 
stress-responsive bacteria, we examine how purely mechanical interactions determine 
colony morphology and internal organization. Growth-induced extensile stresses compete 
with steric constraints, giving rise to the spontaneous formation of microdomains 
composed of highly aligned cells. We characterize this self-organization through 
the distribution of microdomain areas and a nematic order parameter that quantifies 
colony-wide alignment. Mechanosensitivity does not systematically alter domain 
structure, but increasing substrate friction reduces the mean domain size and 
broadens the diversity of orientations. Shifting the balance toward steric 
interactions, by lengthening the cell division size, slows the relaxation of 
colony shape toward isotropy and broadens the distribution of contact forces, 
producing a slower exponential decay. In dense colonies, strong forces are 
transmitted anisotropically through chains of aligned neighbors within microdomains. 
These findings demonstrate that colony-level morphology and stress organization 
can emerge from local mechanical interactions alone, even without requiring 
biochemical signaling. By linking microscopic force transmission to macroscopic 
growth dynamics, our study provides a physical framework for understanding how 
mechanical interactions shape the self-organization of bacterial communities 
under surface confinement.
\end{abstract}

\maketitle
 
\section*{I. Introduction}
Bacteria employ two primary survival strategies:\ they can disperse as motile, 
self-propelled individuals to explore their surroundings, or attach to surfaces and 
form biofilms---dense, immobilized communities that persist under favorable nutrient 
conditions \cite{Du21,Allen19,Gloag20}. During the early stages of biofilm development, 
individual cells aggregate into microcolonies, where physical interactions among growing 
and dividing cells begin to shape the collective architecture. Even in the absence of 
motility, freely growing bacterial colonies expand through extensile forces generated 
by cell elongation and division. Colony growth on surfaces is governed by the interplay 
between adhesive interactions with the substrate and mechanical interactions arising 
from bacterial proliferation and steric constraints. These coupled interactions lead 
to the spontaneous formation of locally aligned regions, or microdomains, whose geometry 
reflects the balance between growth-induced extensile stresses, steric repulsion, and 
interactions with the surrounding environment. Understanding the spatial organization 
of bacterial colonies plays a central role in medicine, biology, and technology, and 
has consequently attracted significant attention \cite{Gloag20,Fei20,Rahbar25,MartinezCalvo22,
Dhar22,Kannan25,DellArciprete18,You21,Farrell13,Zhang21,Amchin22,Cho07,Shao17,Duvernoy18,
Bhusari22,Peterson15}. While colonies tend to appear globally isotropic due to fluctuations 
and imperfect alignment following cell division, local ordering evolves dynamically under 
the influence of key cellular parameters---such as growth rate, division length, and 
stiffness---as well as cell-environment interactions. These environmental effects may 
arise from temporal or spatial variations in nutrient availability, viscoelasticity, 
or temperature, and from confinement and boundary effects. Particularly, cell-surface 
interactions represent a vital yet comparatively underexplored factor in determining 
bacterial colony morphology and internal organization.

The arrangement of bacteria into self-organized ordered microdomains differs from the 
heterogeneous contact structure of dense systems of elongated passive particles. This 
suggests that the pattern of stress transmission through the dynamic network of intercellular 
contacts in bacterial colonies also differs from the force networks observed in static 
granular packings \cite{Majmudar05,Boberski13,Radjai96,Shaebani09b}, even in assemblies 
of elongated particles \cite{Azema10,Azema12}. In passive granular systems, large forces 
are predominantly transmitted along chains that become increasingly guided by cap-to-side 
contacts as particle elongation increases. This raises the question of how strong forces 
are transmitted within bacterial colonies---systems characterized by highly ordered local 
structures and continuous remodeling of the contact network through growth, division, 
and active stress generation. Such dynamic restructuring reflects the inherently active 
nature of bacterial collectives, in which internal growth processes simultaneously 
generate and relax mechanical stresses---an ability with no analogue in passive granular 
assemblies. As a result of the interplay between growth-driven extensile stresses and 
steric contact forces, bacterial colonies develop both active and passive components 
of stress, oriented parallel and perpendicular to the cell elongation axis, respectively 
\cite{Isensee22,You18}. These competing stresses promote the self-organization of bacteria 
into ordered microdomains in two dimensions \cite{Isensee22,You18,Boyer11,Volfson08,
Ghosh15} and can ultimately induce buckling instabilities that trigger the transition 
from two-dimensional to three-dimensional structures \cite{Beroz18,Grant14}.

An additional complexity in stress transmission within bacterial colonies arises from the 
fact that cell growth itself is sensitive to mechanical stress. Experimental studies have 
shown that mechanical forces acting on individual bacteria can modulate their instantaneous 
growth rate \cite{Tuson12,Si15,Wittmann23,Harper23}. In particular, compressive forces applied 
along the major axis of the cell can slow elongation, whereas growth tends to be less sensitive 
to transverse compression except under extremely large forces, where it may cease entirely 
\cite{Si15}. This mechanosensitivity establishes a feedback loop between growth and stress:\ 
local stresses alter growth rates, and the resulting growth in turn reshapes the stress field. 
Consequently, the diversity in cell lengths and orientations that emerges across the colony 
reflects not only steric and frictional interactions but also the local mechanical environment. 
Through this stress-growth coupling, stress-responsive bacteria can adapt their proliferation 
dynamics to mechanical constraints imposed by their surroundings. It remains unclear to what 
extent mechanosensitivity influences the structure and morphology of microdomains in growing 
colonies.

In this work, we investigate how purely mechanical interactions govern the evolution of microdomains 
in growing bacterial colonies confined to surfaces. Using overdamped dynamics simulations of 
nonmotile, stress-responsive bacteria, we vary substrate friction, mechanosensitivity, and 
cell division length to probe their influence on colony morphology, microdomain formation, 
and force transmission. By isolating mechanical effects from biochemical signaling or motility, 
we directly assess whether mechanosensitivity can modify the geometry and alignment of microdomains 
that arise from growth-induced extensile stresses and steric interactions. Our analysis of nematic 
order and contact force distributions reveals how local growth and frictional constraints shape 
the collective organization of the colony. Our study provides a physical framework for understanding 
how bacterial colonies self-organize on surfaces through mechanical feedback alone, highlighting 
the key role of stress-growth coupling in determining microbial community architecture under 
confinement.

\section*{II. Method}
\begin{figure*}
\centering
\includegraphics[width=0.93\textwidth]{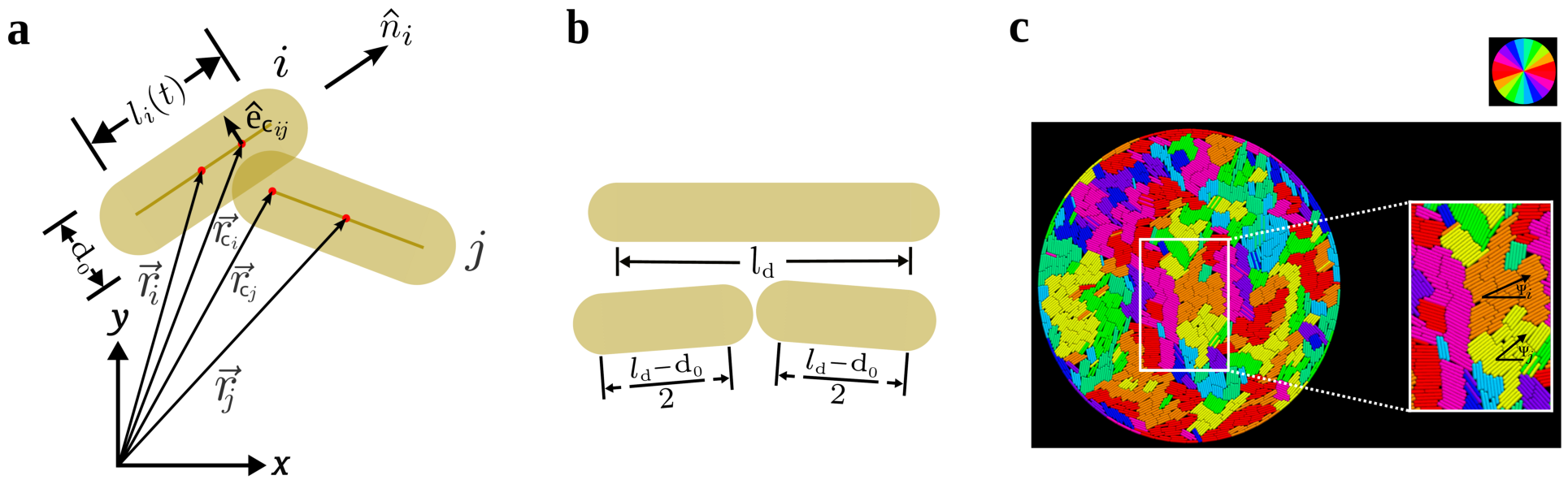}
\caption{(a) Schematic of capsule geometry and cell-cell contact. (b) Cell division into two 
daughter cells when the division length $l_{_\text{d}}$ is reached. (c) Example of a final 
colony configuration showing two microdomains $i$ and $j$ with mean orientations $\Psi_i$ 
and $\Psi_j$ relative to the $x$ axis. Colors correspond to domain orientation as indicated 
by the color wheel.}
\label{Fig:1}
\end{figure*}

In our model, each bacterium is represented as a capsule-shaped particle of fixed diameter $d_{_0}
\,{=}\,0.5\,\mu\text{m}$ and a time-dependent rectangular body length $l(t)$, excluding the two 
hemispherical caps; see Fig.\,\ref{Fig:1}(a). The orientation of the $i$th bacterium is given by 
the unit vector ${\hat n}_{_i}$ along its major axis, with Cartesian components ${\hat n}_{_i} 
\,{=}\, \cos\theta\!_{_i} \hat{x} \,{+}\, \sin\theta\!_{_i} \hat{y}$, where $\theta\!_{_i}$ 
denotes the angle of the cell with respect to the $x$-axis in the laboratory frame. Bacteria 
are confined to a two-dimensional circular domain of radius $R\,{=}\,60\,\mu\text{m}$ and 
interact through repulsive Hertzian contact forces. The force exerted by cell $j$ on cell 
$i$ is given by $\fijvec\,{=}\,E\,d_{_0}^{1{/}2}\,\hij^{3{/}2}\,\ecijhat$, where $E$ is the 
Young's modulus, $\hij$ the overlap between the two interacting cells, and $\ecijhat$ the 
unit vector along the line connecting the nearest points on the axes of the $j$th and $i$th 
cells [Fig.\,\ref{Fig:1}(a)]. The simulation is initiated with a single cell of random position 
and orientation inside the confining boundary. The mechanical stiffness of the cell-wall 
contacts is set to $E_{\text{wall}}\,{=}\,1000\,\text{kPa}$, exceeding the cell-cell Young's 
modulus $E\,{=}\,400\,\text{kPa}$.

In the overdamped limit, the translational dynamics of the $i$th bacterium is governed by 
\cite{You18,Ghosh15,You21}:\
\begin{equation}
\displaystyle\frac{\text{d}{\vec r}\!_{_i}}{\text{d}t}=\displaystyle
\frac{1}{\zeta\,l_i} \sum_{j{=}1}^{N_{\!c}^i} \fijvec, 
\label{Eq:Xevolution} 
\end{equation} 
where $\vec r\!_{_i}$ denotes the cell center position, $N_{\!c}^i$ the number of its contacts, 
and $\zeta$ the drag coefficient per unit length. The dimensional analysis of $\zeta$ gives 
$[\zeta]\,{=}\,[\frac{M}{L T}]$, ensuring that the right-hand side of Eq.\,(\ref{Eq:Xevolution}) 
has the dimension of velocity. Note that the parameter $\zeta$ accounts for the viscous drag 
or substrate friction experienced by a moving bacterium, distinct from intercellular frictional 
resistance, which can be additionally incorporated through the contact force model $\fijvec$. 
This expression corresponds to an overdamped Newtonian dynamics appropriate for nonmotile 
objects in viscous environments. The orientation $\theta\!_{_i}$ of each bacterium evolves 
according to \cite{You18,Ghosh15,You21}:\
\begin{equation}
\displaystyle\frac{\text{d}\theta\!_{_i}}{\text{d}t} = 
\displaystyle\frac{12}{\zeta\,l_i^{^3}} \sum_{j{=}1}^{N_{\!c}^i} 
\big((\vec r\!_{c_{i}}{-}\vec r\!_{_{i}}) {\times} \fijvec \big) 
\,{\cdot}\, \hat z, 
\label{Eq:Theta}
\end{equation}
where $\vec r\!_{c_{i}}{-}\vec r\!_{_{i}}$ is the vector connecting the center of mass of the 
$i$th cell to the contact point on its major axis. 

As the colony grows, cells come into contact and deform one another, modeled through the 
overlap $\hij$. The elements of the local stress tensor is computed from intercellular 
forces using \cite{You21,Shaebani12,Shaebani12b}:\ $\sigma_{_{\eta,\mu}}\,{=}\,\frac{1}{A}(\frac12
\sum_{i{=}1}^{N}\sum_{j{=}1}^{N_{\!c}^i} r\!_{c_{ij,\eta}} f\!_{_{ij,\mu}}\!)$, where 
$\vec r\!_{c_{ij}}{=}\,\vec r\!_{c_{i}}{-}\,\vec r\!_{c_{j}}$, and $A$ is the total area 
of the colony. The equations of motion (\ref{Eq:Xevolution}) and (\ref{Eq:Theta}) are 
integrated using an implicit first-order Euler scheme. For isotropic drag, a suitable 
time step ensuring smooth relaxation of mechanical stresses is estimated as $\Delta t\,
{\approx}\,\zeta{/}E$.

To simulate growing bacteria, individual cell elongation and division are implemented 
at the beginning of each time step. Each bacterium grows at a rate that depends on both 
stochastic variability and mechanical stress. In the absence of stress, a random growth 
rate $\rgi$ is assigned to each cell $i$, drawn uniformly from the range $\big[\frac
{\rg}{2},\frac{3 \rg}{2}\big]$ with mean $\rg\,{=}\,4\,\mu\text{m}{/}\text{h}$. The 
time evolution of the cell length $l_i(t)$ is described by
\begin{align}
\frac{\text{d}l_i(t)}{\text{d}t} \,{=} \left\lbrace 
\begin{array}{cc}
\rgi \,{-}\, \beta \, |\fpi(t)|, \,\,\,\, 
&\rgi \,{>}\, \beta \, |\fpi(t)|, \\
\\
\hspace{-19mm}0, \,\,\,\, &\rgi \,{\leq}\, \beta \, |\fpi(t)|,
\end{array} \right. 
\label{Eq:Growth}
\end{align}
where $|\fpi(t)|\,{=}\,\sum_{j{=}1}^{N_{\!c}^i}\big|\fijvec(t){\cdot}\,{\hat n}_{_i}\big|$ 
represents the total force projected along the major axis of the cell, and $\beta$ 
quantifies the mechanosensitivity of growth, characterizing the strength of cell 
response to mechanical stimuli. To capture the observed inhibition of growth under 
strong lateral compression, an additional condition is imposed:\ if the overlap 
between a cell and any of its neighbors exceeds $d_{_0}{/}2$, elongation is 
temporarily halted until the overlap decreases below this threshold. This 
constraint overrides Eq.\,(\ref{Eq:Growth}). When a bacterium reaches the division 
length $l_{_\text{d}}$, it divides into two daughter cells; see Fig.\,\ref{Fig:1}(b). 
The daughter cells inherit the orientation of the parent with a small random deviation 
of up to $10^{\circ}$, and their individual growth rates are independently sampled 
from the same uniform distribution $\big[\frac{\rg}{2},\frac{3 \rg}{2}\big]$.

This modeling framework allows us to simulate the collective dynamics of stress-responsive, 
nonmotile bacteria in confined two-dimensional environments, capturing both mechanical 
interactions and growth-driven feedback processes that shape the evolving colony morphology. 
Although exchanging contact-force information across the boundaries of decomposed domains 
presents technical challenges, such dynamic contact networks can be efficiently parallelized 
for large-scale simulations through adaptive hierarchical domain decomposition combined 
with dynamic load balancing \cite{Shojaaee12}.

To quantify orientational heterogeneity within the colony, we partition bacteria into 
microdomains based on local alignment. Each microdomain $i$ is characterized by an 
average orientation $\Psi_i$ measured with respect to the $x$-axis. Two contacting 
bacteria belong to the same microdomain if the relative angle between them is less 
than or equal to $10^{\circ}$. An example of such a decomposition is 
shown in Fig.\,\ref{Fig:1}(c), where different microdomains are indicated by distinct 
colors. To characterize the global orientational order of the colony, we define a 
nematic order parameter $\sigma\,{=}\,\big\langle \cos(2 \theta_{ij})\big\rangle_{ij}\,
{=}\,\big\langle 2 \cos^{2}(\theta_{ij})\,{-}\,1\big\rangle_{ij}$, where $\theta_{ij}\,
{=}\,\theta_j{-}\theta_i$ denotes the relative orientation between bacteria $i$ and 
$j$, and the average is taken over all interacting pairs.

\begin{figure*}[t]
\centering
\includegraphics[width=0.99\textwidth]{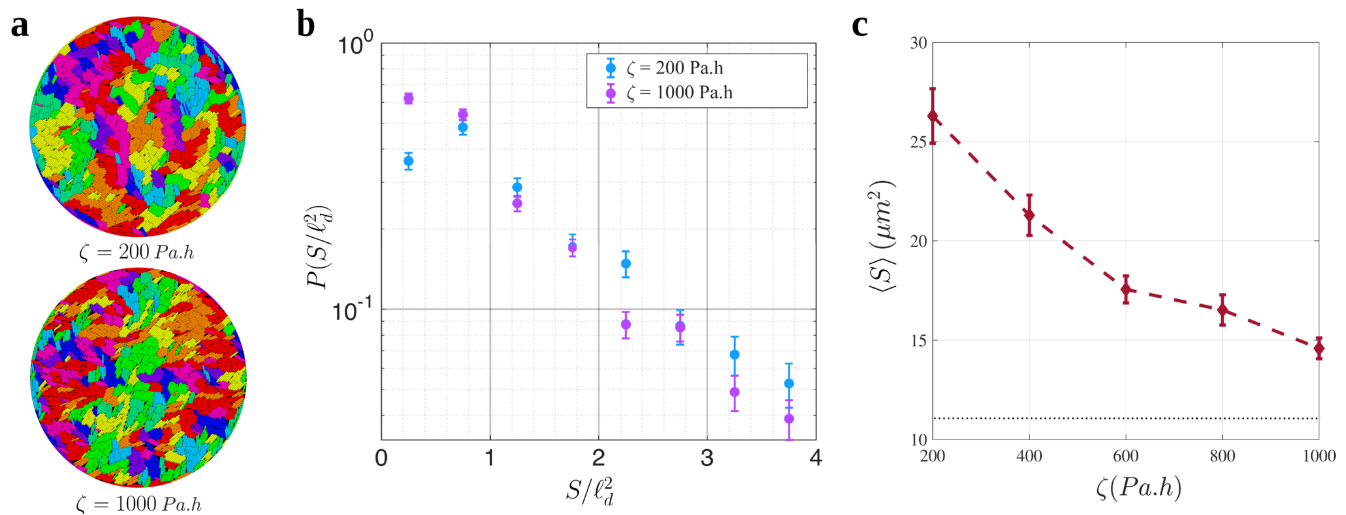}
\caption{(a) Final configurations of bacterial colonies growing within circular confinement 
for two substrate friction coefficients $\zeta\,{=}\,200$ and $1000\,\text{Pa.h}$. Other 
parameters: $l_{_\text{d}}\,{=}\,3\,\mu\text{m}$, $\beta\,{=}\,0.02\,(\mu\text{m.kPa.h})^{-1}$. 
(b) Probability distribution of rescaled microdomain areas for different friction values. (c) 
Mean microdomain area as a function of substrate friction. Error bars indicate the standard 
error of the mean. The dotted line indicates the approximate mean microdomain size in 
frictionless packings of passive particles whose elongation matches the maximum elongation 
of the bacteria; see text.}
\label{Fig:2}
\end{figure*}

\section*{III. Results and Discussion}

Each simulation begins with a single bacterium with an arbitrary orientation at the center 
of the circular domain. As the bacteria proliferate, the colony grows freely until it 
reaches the circular confinement boundary and eventually settles into a stationary 
configuration that fills the confinement. At this stage, proliferation ceases as the 
buildup of mechanical stress suppresses further growth (see Suppl.\ Movie). We begin 
by systematically examining how frictional interactions with the underlying substrate 
influence the emergence of self-organized ordering in expanding surface-confined colonies 
in Subsec.\,3.1. We then investigate the role of mechanosensitivity---i.e., the feedback 
between growth and mechanical stress---in shaping the structure of the microdomains that 
arise during colony expansion in Subsec.\,3.2. Finally, we explore how increasing the 
relative contribution of steric interactions alters both the degree of orientational 
ordering and the transmission of stress across the packing in Subsec.\,3.3. Together, 
these analyses elucidate and disentangle the key physical mechanisms that govern the 
collective organization of densely growing bacterial populations.

\subsection*{3.1 Substrate friction suppresses self-organized ordering}
\label{Sec:Friction}
To examine how friction between bacteria and the underlying substrate influences 
self-organized microdomain formation and the development of local orientational order, 
we perform simulations over a range of values of the cell-substrate friction coefficient 
$\zeta$. More generally, $\zeta$ represents the viscous resistance experienced by a 
bacterium as it moves through the environment. However, because our system is confined 
to two dimensions and motion is limited to interactions with the substrate, $\zeta$ is
most appropriately interpreted here as an effective cell-substrate friction coefficient.

Figure \ref{Fig:2}(a) shows representative stationary configurations of colonies after 
fully filling the circular confinement for low and high values of $\zeta$. At higher 
friction, the colony breaks into noticeably smaller microdomains of aligned bacteria, 
and the orientations of these domains are more broadly distributed. In contrast, lower
friction promotes the formation of larger, more coherently aligned microdomains, 
indicating enhanced local orientational order. These differences arise because the 
tendency of elongated bacteria to align with their neighbors competes with frictional 
resistance from the substrate. At low friction, bacteria can more readily rotate, slide, 
and reorient to align with the local director, whereas at high friction such reorientation 
becomes increasingly costly, leading to reduced alignment and a more disordered local 
structure. For a more quantitative comparison, we compute the probability distribution 
$P(S)$ of microdomain areas $S$ for different values of the substrate friction 
coefficient. Previous work has shown that the tail of $P(S)$ decays exponentially, 
with a slope that decreases as the division length $l_{_\text{d}}$ increases, 
corresponding to larger mean microdomain sizes \cite{You18}. Consistent with 
these findings, we observe that the tails of $P(S)$ exhibit an approximately 
exponential decay across all friction values considered; see e.g.\ Fig.\,\ref{Fig:2}(b). 
Increasing the friction coefficient $\zeta$ leads to a slightly steepening of 
the exponential tail. The corresponding decrease in the mean microdomain area 
is more pronounced, as presented in Fig.\,\ref{Fig:2}(c). Notably, growth-driven 
dynamics prevent the complete disappearance of local orientational order even 
in the large-friction limit. This behavior stands in marked contrast to 
packings of passive elongated particles, in which local alignment is considerably 
weaker \cite{Azema10,Azema12,Rocks23,Marschall18}. For comparison, at high friction 
($\zeta\,{=}\,1000\,\text{Pa.h}$) the mean number of our bacteria with division length 
$l_{_\text{d}}\,{=}\,3\,\mu\text{m}$ within a locally aligned microdomain exceeds 
eight, whereas in frictionless assemblies of passive particles with elongation 
matching the maximum elongation of these bacteria, the mean size of aligned clusters 
does not exceed two to three particles \cite{Rocks23,Marschall18}.

\begin{figure*}[t]
\centering
\includegraphics[width=0.99\textwidth]{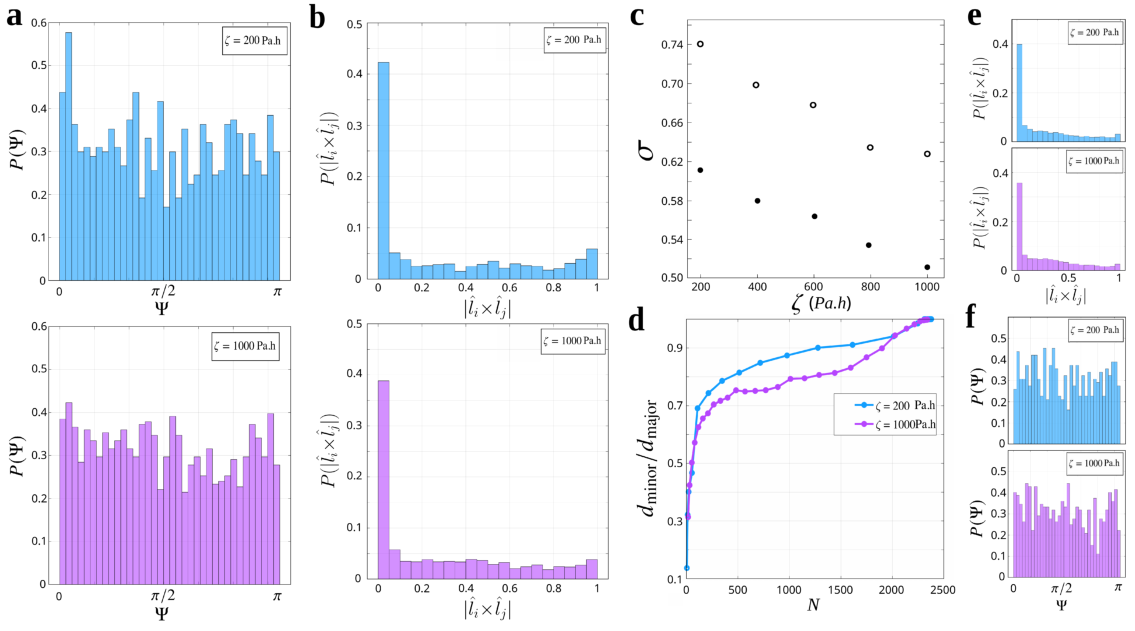}
\caption{(a,b) Probability distribution of domain orientations, $P(\Psi)$, and 
probability distribution of the magnitude of the cross product between the unit 
orientation vectors of contacting bacteria, $P(|\hat l_i{\times}\hat l_j|)$, 
for the final colony configurations and different values of the substrate 
friction coefficient $\zeta$. Lower friction promotes stronger local alignment, 
whereas higher friction leads to a more isotropic orientation distribution. 
Other parameters: $l_{_\text{d}}\,{=}\,3\,\mu\text{m}$, $\beta\,{=}\,0.02\,
(\mu\text{m.kPa.h})^{-1}$. (c) Nematic order parameter versus substrate friction 
coefficient for the final configurations (full circles) and freely growing 
colonies before reaching the boundaries (open circles). (d) Evolution of the 
minor-to-major axis ratio of the colony as a function of bacterial population 
size for both friction coefficients. (e,f) $P(|\hat l_i{\times}\hat l_j|)$ 
and $P(\Psi)$ for freely growing colonies and different values of $\zeta$.}
\label{Fig:3}
\end{figure*}

Next, we investigate how the increase in the number of microdomains with friction 
affects the diversity of their orientations and the resulting collective alignment 
of the colony. Figure \ref{Fig:3}(a) shows the probability distribution of microdomain 
orientations, $P(\Psi)$, for low and high substrate friction coefficients. At higher 
friction, the distribution becomes visibly more uniform, indicating a broader spread 
of domain orientations. For a quantitative comparison, the mean and standard deviation 
of $P(\Psi)$ are $\text{m}\,{=}\,0.3157$ and $\text{std}\,{=}\,0.0798$ for $\zeta
\,{=}\,200\,\text{Pa.h}$, and $\text{m}\,{=}\,0.3157$ and $\text{std}\,{=}\,0.0574$ 
for $\zeta\,{=}\,1000\,\text{Pa.h}$. The relative fluctuation, defined as $\eta\,{=}
\,\frac{\text{std}}{\text{m}}$, therefore decreases from $\eta\,{\approx}\,0.25$ at 
low friction to $\eta\,{\approx}\,0.17$ at high friction; a factor of 1.5 difference. 
This indicates that the microdomain orientations are significantly more diverse at 
higher frictions.

To quantify the overall orientational order of the colony, we consider orientations 
at single-contact level and first examine the magnitude of the cross product between 
the unit orientation vectors of contacting bacteria, $|\hat l_i{\times}\hat l_j|$. 
This measure ranges from 0 for parallel neighbors to 1 for perpendicular contacts. 
The colony-averaged value $\langle|\hat l_i{\times}\hat l_j|\rangle$ is approximately 
0.28 for $\zeta\,{=}\,200\,\text{Pa.h}$ and 0.29 for $\zeta\,{=}\,1000\,\text{Pa.h}$, 
corresponding to a modest difference of about $3\%$ and indicating enhanced alignment 
at lower friction. A comparison of the full probability distributions $P(|\hat l_i{
\times}\hat l_j|)$ in Fig.\,\ref{Fig:3}(b) reveals that, at low friction, the distribution 
develops more pronounced peaks near both extremes (i.e., close to 0 and 1). This effect 
increases the standard deviation from $0.0786$ at $\zeta\,{=}\,1000\,\text{Pa.h}$ 
to $0.0858$ at $\zeta\,{=}\,200\,\text{Pa.h}$, representing an increase of nearly 
$9\%$. The sharpening of the peaks at low friction reflects two concurrent effects:\ 
enhanced microdomain formation, which increases the number of nearly perfectly 
aligned contacting pairs, and stronger orientational mismatches at microdomain 
boundaries, where contacts between neighboring domains produce larger relative 
angles. While the quantity $|\hat l_i{\times}\hat l_j|$ provides useful information 
about overall alignment, we find the nematic order parameter $\sigma$ to be a more 
robust global measure, as it more clearly discriminates between different frictional 
regimes. As shown in Fig.\,\ref{Fig:3}(c), $\sigma$ decreases by nearly $20\%$ when 
the friction coefficient increases from $\zeta\,{=}\,200\,\text{Pa.h}$ to $\zeta\,
{=}\,1000\,\text{Pa.h}$.

\begin{figure*}
\centering
\includegraphics[width=0.99\textwidth]{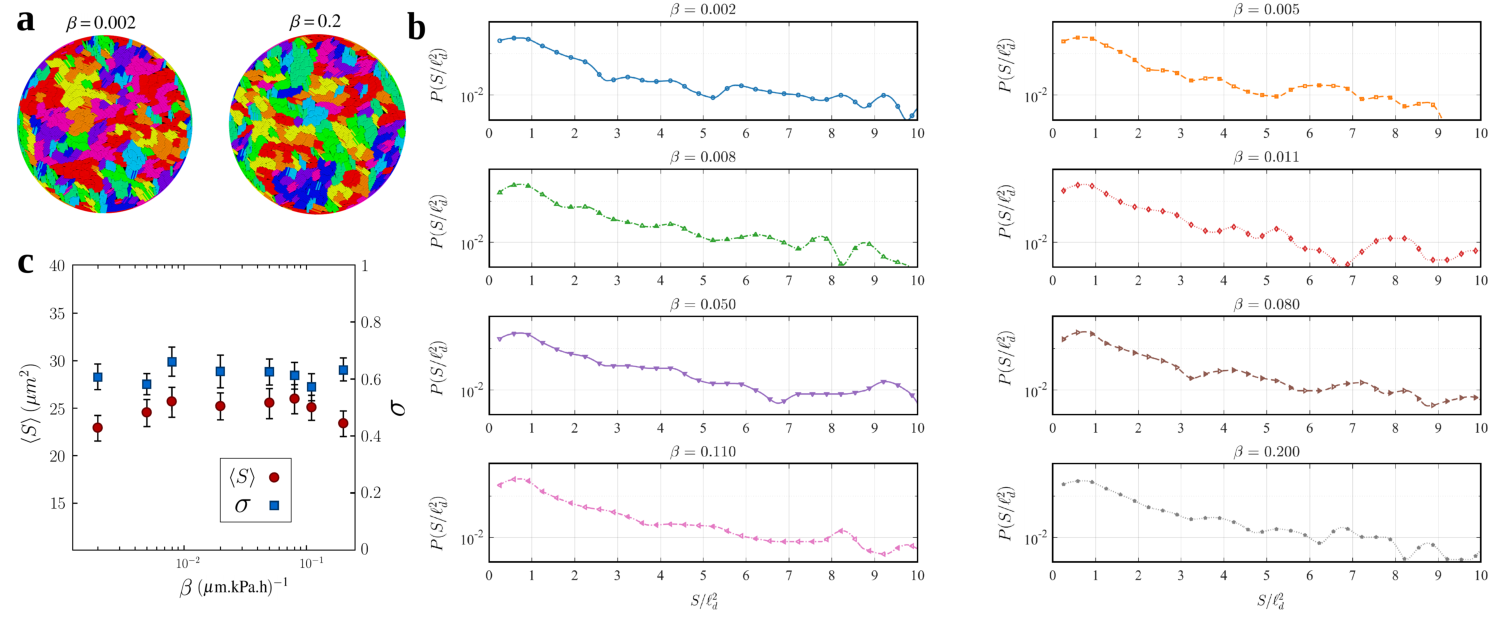}
\caption{(a) Final configurations of bacterial colonies growing under circular 
confinement for two different values of mechanosensitivity $\beta\,{=}\,0.002$ 
and $0.2\,(\mu\text{m.kPa.h})^{-1}$. Other parameters: $l_{_\text{d}}\,{=}\,3\,
\mu\text{m}$, $\zeta\,{=}\,200\,\text{Pa.h}$. (b) Probability distribution of 
microdomain areas for different values of $\beta$. (c) Mean microdomain area 
$\langle S \rangle$ and global nematic order parameter $\sigma$ vs $\beta$.}
\label{Fig:4}
\end{figure*}

The friction-dependent ability of bacteria to rotate, slide, and align with their 
neighbors has direct consequences for the morphology of the growing colony on the 
surface. We quantify colony shape anisotropy using the ratio of the minor to major 
diameters of the evolving colony, $d_\text{minor}\,{/}\,d_\text{major}$, which 
ranges from 0 for a highly elongated colony to 1 for a perfectly circular one. 
As shown in Fig.\,\ref{Fig:3}(d), lower friction initially favors the development 
of a strongly elongated colony along the orientation of the founding bacterium, 
resulting in smaller values of $d_\text{minor}\,{/}\,d_\text{major}$ compared 
to higher friction. Over time, however, the enhanced rotational mobility at low 
friction, together with stochastic growth and division, promotes the formation 
of multiple microdomains with diverse orientations. This progressive reorientation 
drives the colony morphology toward a more isotropic shape, causing $d_\text{minor}
\,{/}\,d_\text{major}$ at low friction to rapidly exceed that observed at higher 
friction. This trend persists until the expanding colonies encounter the circular 
confinement, at which point the rigid boundary enforces a circular morphology in 
both cases.

After the growing colony first encounters the rigid circular boundary while 
still exhibiting an anisotropic shape, interactions with the confinement 
begin to influence the formation and orientation of newly generated 
microdomains. Previously formed microdomains in the colony interior 
may also be affected by this boundary-induced reorganization. Growth 
ultimately ceases once the circular domain becomes fully occupied by 
bacteria. This raises the question of whether the influence of substrate 
friction on the development of nematic order differs between the final, 
mechanically arrested state of the colony (discussed above) and earlier 
stages of growth, during which the colony expands freely without boundary 
interactions. To address this question, Figs.\,\ref{Fig:3}(e,f) show the 
probability distributions $P(|\hat l_i{\times}\hat l_j|)$ and $P(\Psi)$ 
for low and high friction coefficients at comparable time points during 
the freely growing phase. During this stage, differences in the relative 
fluctuations of $P(\Psi)$ are less pronounced, with values of approximately 
0.24 and 0.26 for decreasing friction. However, the statistics of the 
local alignment measure $|\hat l_i{\times}\hat l_j|$ exhibit strong trends 
similar to those observed in the final static configurations. Specifically, 
the colony-averaged value of $\langle|\hat l_i{\times}\hat l_j|\rangle$ 
is approximately 0.26 for $\zeta\,{=}\,200\,\text{Pa.h}$ and 0.27 for 
$\zeta\,{=}\,1000\,\text{Pa.h}$, corresponding to a difference of 
nearly $4\%$. Moreover, the standard deviation of $P(|\hat l_i{\times}
\hat l_j|)$ increases from 0.0741 at high friction to 0.0829 at low 
friction, representing an enhancement of nearly $12\%$. Consistent 
with these observations, the nematic order parameter $\sigma$ also 
decreases markedly with increasing friction, exhibiting a reduction 
of nearly $17\%$ when $\zeta$ is increased from $200\,\text{Pa.h}$ 
to $1000\,\text{Pa.h}$; see Fig.\,\ref{Fig:3}(c). These results indicate 
that the friction-dependent suppression of orientational order is already 
established during the freely growing phase and is subsequently modulated, 
but not qualitatively altered, by confinement effects.

\subsection*{3.2 Mechanosensitive growth weakly modulates local ordering}
\label{Sec:Mechanosensitivity}

The degree of bacterial mechanosensitivity, $\beta$, introduced in Eq.\,\ref{Eq:Growth}, 
quantifies the strength of the feedback between evolving internal stresses and cellular 
growth dynamics. In previous work, we investigated the growth of three-dimensional 
colonies of stress-responsive bacteria under isotropic confining pressure \cite{Rahbar25}. 
We showed that, at finite confining pressure, increasing $\beta$ leads to a measurable, 
though modest, reduction in bacterial population size and colony volume, accompanied by
an increase in the doubling time of bacteria. By contrast, as the imposed pressure is 
reduced and the colony approaches a freely growing regime, the influence of mechanosensitivity 
diminishes and the expected exponential population growth associated with unconstrained 
expansion is recovered.

This raises the question of whether mechanosensitivity influences self-organized ordering 
in bacterial colonies growing on surfaces. Even if such an effect exists, one expects 
it to be moderate during freely growing phases, since stresses generated by growth 
and proliferation can be readily relaxed through colony expansion. The impact of 
mechanosensitivity may therefore become more apparent once the colony experiences 
the circular confinement. To assess this possibility, we compare the final static 
configurations of surface-confined colonies generated for different values of $\beta$. 
We systematically vary $\beta$ over several orders of magnitude and analyze the 
resulting steady-state colony structures. Representative configurations for $\beta
\,{=}\,0.002$ and $0.2\,(\mu\text{m.kPa.h})^{-1}$ are shown in Fig.\,\ref{Fig:4}(a). 
The resulting microdomains exhibit similar morphologies in terms of both area and 
orientational diversity.

To quantify these observations, we examine the probability distribution $P(S)$ of 
microdomain areas $S$ for different levels of mechanosensitivity. As shown in 
Fig.\,\ref{Fig:4}(b), the distributions display a pronounced peak and an approximately 
exponential decay for all values of $\beta$, with no systematic dependence of the 
tail slope on mechanosensitivity. Consistently, Fig.\,\ref{Fig:4}(c) shows that 
the mean microdomain area remains essentially unchanged across the explored range 
of $\beta$. We further compute the nematic order parameter $\sigma$ and find 
no clear trend with mechanosensitivity when varying $\beta$ by several orders 
of magnitude.

To conclude, our results indicate that, within the biologically relevant range 
of growth rates, stress-responsive growth does not play a decisive role in the 
formation or restructuring of microdomains in surface-confined bacterial colonies. 
Because growth-induced stresses can be efficiently relaxed through expansion prior 
to mechanical arrest, the effects of mechanosensitivity remain too weak to 
significantly alter self-organized ordering.

\subsection*{3.3 Role of proliferation dynamics in colony ordering and stress transmission}
\label{Sec:Proliferation}

\begin{figure}[t]
\centering
\includegraphics[width=0.47\textwidth]{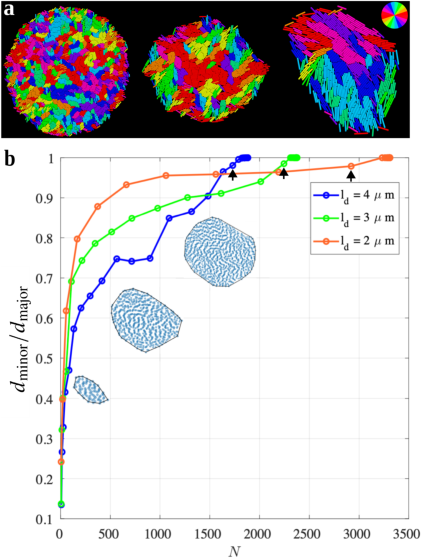}
\caption{(a) Simulated configurations at concurrent time steps for $\zeta\,{=}\,200
\,\text{Pa.h}$, $\beta\,{=}\,0.02\,(\mu\text{m.kPa.h})^{-1}$, and different division 
lengths $l_{_\text{d}}\,{=}\,2\,\mu\text{m}$ (left), $3\,\mu\text{m}$ (middle), and 
$4\,\mu\text{m}$ (right). The corresponding number of bacteria are $N\,{=}\,2853$, 
$1265$, and $505$, respectively. Color circle shows the orientational distribution. 
(b) Evolution of the minor-to-major axis ratio of the bacterial colony as a function 
of the number of bacteria within the colony for different division lengths. Insets 
show how the global shape anisotropy of the colony evolves towards the final isotropic 
one. Black arrows mark the moment when the expanding colony reaches the confining 
walls.}
\label{Fig:5}
\end{figure}

Beyond environmental interactions such as substrate friction, the intrinsic growth and 
proliferation dynamics of bacteria constitute a central control mechanism for self-organized 
ordering and morphological evolution in surface-confined colonies. In our simulations, 
the intrinsic growth rate $\rg$ is fixed within a biologically relevant range. Variations 
of $\rg$ within this range do not qualitatively alter the emergent ordering (though broader 
changes of $\rg$ beyond this range was reported to affect the mean area of microdomains \cite{You18}). 
Instead, we focus on the division length $l_d$, which directly controls the typical aspect 
ratio of cells at division and thereby tunes the relative importance of steric interactions 
during growth. It has been shown that increasing $l_d$ enhances the nematic order across 
the colony and decouples the active and passive stress components \cite{Isensee22,You18}. 

Figure \ref{Fig:5}(a) shows representative colony configurations at concurrent times for 
three different division lengths, $l_{_\text{d}}\,{=}\,2,\,3,$ and $4\,\mu\text{m}$, at 
fixed substrate friction and mechanosensitivity. Increasing the division length reduces 
the total number of bacteria required to fill the confinement, leading to colonies composed 
of fewer but more elongated cells. These colonies exhibit visibly enhanced local alignment 
and reduced orientational disorder as $l_d$ increases, consistent with previous reports. 
The color-coded configurations reveal that larger division lengths promote the formation 
of more coherent and extended microdomains.

\begin{figure*}
\centering
\includegraphics[width=0.85\textwidth]{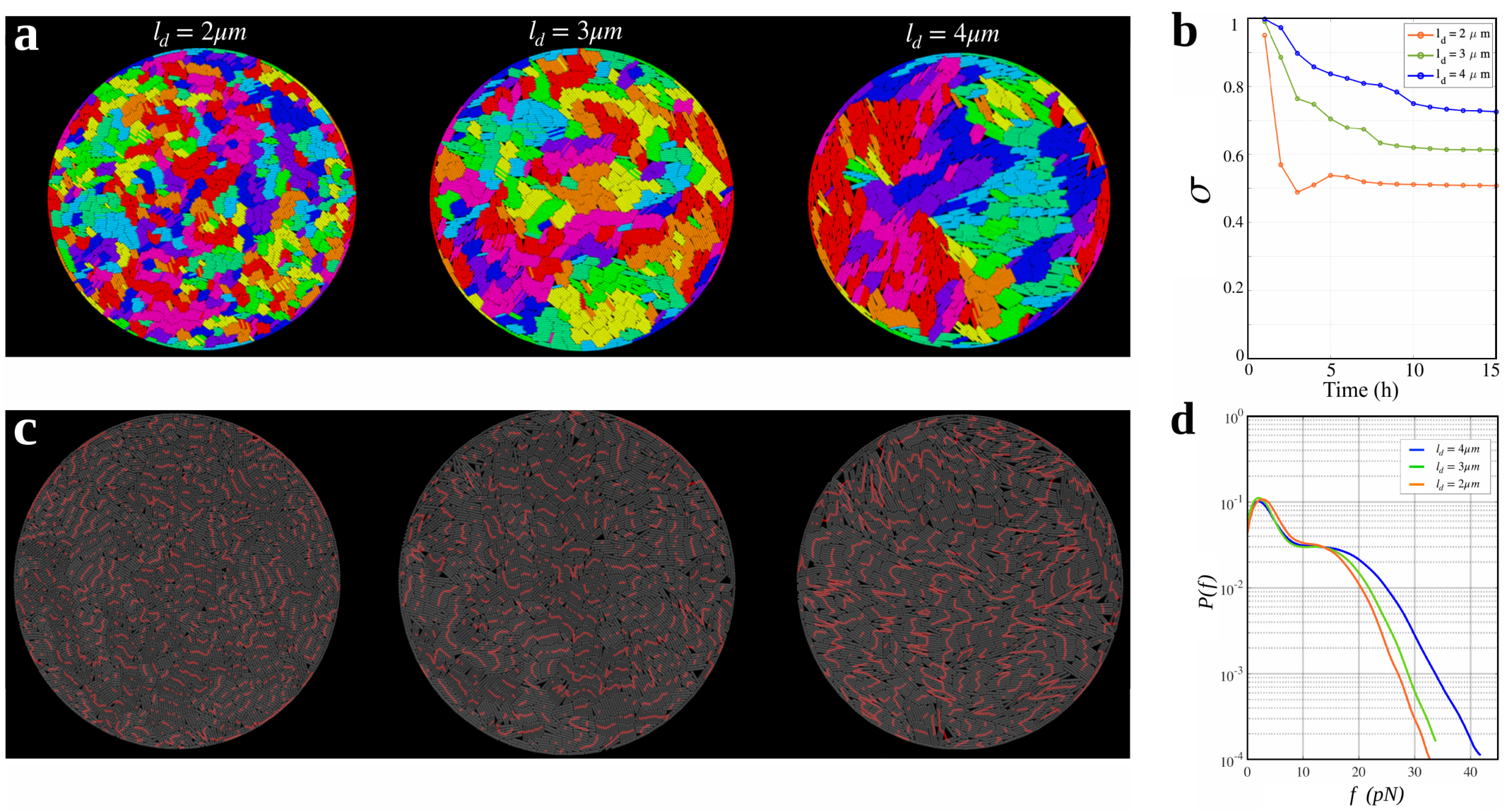}
\caption{(a) Stationary configurations of bacterial colonies after the circular confinement 
becomes fully populated, shown for a fixed substrate friction coefficient $\zeta\,{=}\,200
\,\text{Pa.h}$, mechanosensitivity $\beta\,{=}\,0.02\,(\mu\text{m.kPa.h})^{-1}$, and varying 
division lengths $l_{_\text{d}}$. Cell colors indicate local orientations. (b) Time evolution 
of the nematic order parameter for different division lengths $l_{_\text{d}}$. (c) Force 
networks (red lines) formed by contacts with magnitudes exceeding $|f|\,{=}\,13\,\text{pN}$, 
overlaid on the corresponding colony configurations shown in panel (a). (d) Probability 
distributions of individual intercellular contact forces for different division lengths 
$l_{_\text{d}}$.}
\label{Fig:6}
\end{figure*}

The proliferation dynamics also strongly influence the temporal evolution of colony shape. 
To quantify this effect, we monitor the ratio of the minor to major colony diameters, 
$d_\text{minor}\,{/}\,d_\text{major}$, as a function of the bacterial population size; 
see Fig.\,\ref{Fig:5}(b). Colonies initially grow anisotropically, reflecting the 
directional bias inherited from the original bacterium. This anisotropy persists longer 
for larger division lengths, indicating that colonies composed of more elongated bacteria 
relax their shape more slowly. The slower relaxation can be attributed to the smaller 
population size and reduced frequency of division-induced rearrangements, rendering 
these colonies mechanically stiffer and more resistant to orientational restructuring. 
Once the expanding colonies reach the rigid circular boundary (marked by black arrows), 
the imposed confinement suppresses further anisotropic growth and drives all colonies 
toward an isotropic final shape, as shown in Fig.\,\ref{Fig:6}(a).

The evolution of orientational order further highlights the role of proliferation 
dynamics. Figure \ref{Fig:6}(b) shows the time evolution of the nematic order 
parameter $\sigma$ for different division lengths. For all cases, $\sigma$ gradually 
decreases during growth due to proliferation-induced rearrangements. However, this decay 
is significantly slower for larger $l_d$ and the final stationary value of $\sigma$ 
is systematically higher (nearly a linear increase with $l_d$). The stationary state 
is reached after approximately $5.5$ hours for $l_{_\text{d}}\,{=}\,2\,\mu\text{m}$, 
$7.5$ hours for $l_{_\text{d}}\,{=}\,3\,\mu\text{m}$, and $10$ hours for $l_{_\text{d}}
\,{=}\,4\,\mu\text{m}$. Our results thus show that colonies composed of more elongated 
bacteria retain a higher degree of nematic order throughout their evolution and in 
the final confined state.

These differences in ordering are closely linked to the transmission of mechanical 
stresses within the colony. Figure \ref{Fig:6}(c) visualizes the network of strong 
intercellular forces, highlighting contact forces with magnitudes exceeding $|F|\,
{=}\,13\,\text{pN}$. As the division length increases, strong forces increasingly 
organize into extended chains running predominantly along side-to-side contacts 
between neighboring bacteria. This trend is quantitatively confirmed by the probability 
distributions of individual contact forces shown in Fig.\,\ref{Fig:6}(d). While all 
distributions exhibit an approximately exponential tail, the probability of encountering 
large forces grows with increasing $l_d$, reflecting the enhanced steric coupling 
associated with longer cells.

Importantly, this behavior differs qualitatively from force transmission in packings 
of passive elongated granular particles. In passive systems, increasing particle 
elongation typically shifts the dominant contribution of large forces from cap-to-cap 
contacts toward cap-to-side contacts \cite{Azema10,Marschall18}. In contrast, in proliferating 
bacterial colonies, growth-driven extensile stresses favor the emergence of side-to-side 
force chains within aligned microdomains. These force networks are continuously remodeled 
by growth and division, producing transient yet highly anisotropic stress pathways that 
have no direct analogue in static passive granular assemblies. Therefore, our results 
demonstrate that bacterial proliferation dynamics, specifically the division length, 
play a decisive role in governing colony ordering, shape evolution, and stress 
transmission. While environmental interactions set important constraints, the coupling 
between growth, division, and steric interactions provides an intrinsic mechanism by 
which bacterial colonies regulate their internal organization and mechanical architecture 
under surface confinement.

\section*{IV. Conclusion and Outlook}

In this work, we have shown that the collective organization of bacterial colonies growing 
on surfaces can emerge purely from mechanical interactions generated by growth and steric 
constraints. Using overdamped dynamics simulations of nonmotile, stress-responsive bacteria, 
we demonstrated that growth-induced extensile stresses, combined with cell-cell and 
cell-substrate interactions, are sufficient to drive the spontaneous formation of 
aligned microdomains and anisotropic stress networks, even in the absence of biochemical 
signaling or motility.

By systematically varying substrate friction, mechanosensitivity, and division length, we 
disentangled the relative roles of environmental interactions and intrinsic proliferation 
dynamics. Substrate friction was found to be a key external control parameter:\ increasing 
friction suppresses local reorientation, reduces microdomain size, and leads to a measurable 
decrease in nematic order. In contrast, mechanosensitive growth, despite introducing a 
feedback between stress and proliferation, does not qualitatively alter microdomain 
structure or global ordering within the biologically relevant range of growth rates. 
This robustness reflects the ability of growing colonies to efficiently relax internally 
generated stresses through expansion prior to mechanical arrest. Crucially, we identified 
bacterial proliferation dynamics as an intrinsic and decisive factor governing colony 
ordering and stress transmission. While the intrinsic growth rate itself plays a minor 
role, the division length, which controls the effective aspect ratio of cells, strongly 
influences colony morphology, the relaxation of shape anisotropy, and the organization 
of force networks. Colonies composed of more elongated cells evolve more slowly toward 
isotropy, sustain higher nematic order, and develop force chains dominated by 
side-to-side contacts within aligned microdomains. This mode of stress transmission stands 
in sharp contrast to passive granular packings of elongated particles, where increasing 
elongation shifts strong forces toward cap-to-side contacts. The difference highlights 
the fundamentally active nature of bacterial colonies, in which growth continuously 
generates, redistributes, and relaxes mechanical stresses.

Our results establish a physical framework in which colony-scale morphology and internal 
organization arise from the coupling between growth, steric interactions, and mechanical 
constraints imposed by the environment. This framework provides a unifying explanation 
for how ordered microstructures and anisotropic stress pathways emerge in dense bacterial 
assemblies without invoking biochemical coordination.

Looking forward, several extensions of this work merit exploration. Incorporating 
heterogeneous substrates, spatially varying friction, or viscoelastic environments 
would allow closer connections to biologically realistic settings. While circular 
confinement was adopted here to avoid boundary-induced bias in self-organized 
ordering, the numerical framework can be extended to examine how colony organization 
can be altered or tuned by confinement geometry or by applying a finite confining pressure. 
Building on our previous methods for isotropic compression of passive granular 
assemblies \cite{Shaebani09,Sadjadi08} and growing bacterial colonies \cite{Rahbar25}, 
such approaches would enable direct modeling of growth under soft constraints, 
such as confinement by agarose pads \cite{Grant14} or finite (hydrostatic) pressure 
\cite{Kumar13,Sinha17,Mota18,Nepal18}. More broadly, comparing the mechanical 
response and stress-induced rearrangements of passive systems \cite{Shaebani07,
Goldenberg05,Shaebani08,Kolb06,Ostojic06} with those of actively growing colonies 
may clarify how growth-driven stress generation fundamentally alters force 
transmission and collective dynamics. These directions highlight the potential 
role of mechanical constraints not merely as passive boundary conditions, but 
as active regulators of collective behavior in growing biological matter.

\section*{Acknowledgments}

This work was supported by the Deutsche Forschungsgemeinschaft 
(DFG) within the collaborative research center SFB 1027 and also 
via grants INST 256/539-1, which funded the computing resources 
at Saarland University. R.S.\ acknowledges support by the Young 
Investigator Grant of Saarland University, Grant No.\ 7410110401. 
ChatGPT (OpenAI) was used to enhance readability.

\bibliography{Refs-Bacteria}

\end{document}